%% file: writeup.tex
\documentclass[11pt]{article} 
\usepackage{fullpage}
\usepackage{amsmath}
\usepackage{amssymb}
\usepackage{amsthm}
\usepackage{graphicx}
\usepackage{enumerate}
\usepackage{algorithm}
\usepackage[noend]{algpseudocode}
\newenvironment{varalgorithm}[1]
  {\algorithm\renewcommand{\thealgorithm}{\textnormal{#1}}}
  {\endalgorithm}
\usepackage[usenames,dvipsnames]{color}

\newtheorem{lemma}{Lemma}

\newtheorem*{theorem1}{Theorem 1}
\newtheorem*{theorem2}{Theorem 2}
\newtheorem*{theorem3}{Theorem 3}
\newtheorem*{theorem4}{Theorem 4}

\newcommand{\E}{\mathrm{E}}

\title{Online Learning and Profit Maximization\\ from Revealed Preferences}
\author{
Kareem Amin\footnotemark[1], Rachel Cummings\footnotemark[2], Lili Dworkin\footnotemark[1], Michael Kearns\footnotemark[1], Aaron Roth\footnotemark[1]
}
\date{}

\begin{document}
\maketitle
\renewcommand{\thefootnote}{\fnsymbol{footnote}}
\footnotetext[1]{Computer and Information Science, University of Pennsylvania;
\texttt{\{akareem,ldworkin,mkearns,aaroth\}@cis.upenn.edu}
}
\footnotetext[2]{Computing and Mathematical Sciences, California Institute of Technology;
\texttt{rachelc@caltech.edu}. Research performed while the author was visiting the
University of Pennsylvania.
}
\renewcommand{\thefootnote}{\arabic{footnote}}
\begin{abstract}
We consider the problem of learning from revealed preferences in an online setting. In our framework, each
period a consumer buys an optimal bundle of goods from a merchant according to her (linear) utility function and current
prices, subject to a budget constraint. The merchant observes only the purchased goods, and seeks to adapt prices to
optimize his profits. We give an efficient algorithm for the merchant's problem that consists of a learning phase
in which the consumer's utility function is (perhaps partially) inferred, followed by a price optimization step.
We also consider an alternative online learning algorithm for the setting where prices are set exogenously, but the
merchant would still like to predict the bundle that will be bought by the consumer for purposes of inventory or supply chain
management. 

In contrast with most prior work on the revealed preferences problem, we demonstrate that by making stronger assumptions
on the form of utility functions, efficient algorithms for both learning and profit maximization are possible,
even in adaptive, online settings.
\end{abstract}

\thispagestyle{empty}
\newpage
\clearpage
\setcounter{page}{1}

\section{Introduction}
\input{introduction}

\section{Preliminaries}
\input{preliminaries}

\section{Maximizing Profit in the Price-Setting Model}

We begin by considering the first model, in which the merchant controls prices, and seeks to maximize profit. First we show that, given the coefficients $v_i$ of the consumer's linear utility function,
we can efficiently compute the profit-maximizing prices. We will then combine this algorithm
with a query algorithm for learning the coefficients, thus yielding an online no-regret pricing algorithm.

\subsection{Computing Optimal Prices Offline}
\input{optprices}

\subsection{Learning Consumer Valuations}
\input{learnval}

\subsection{Putting It All Together}
\input{puttogether}

\section{Predicting Bundles in the Exogenous Price Model}

\input{exogenous}

\section{Conclusion and Future Work}
Our work provides efficient learning algorithms for two new preference learning models. We leave as an open question whether it is possible to remove our assumptions of discretized and lower-bounded valuations. In the price-setting model, one might also wish to devise an algorithm in which the merchant can approximately optimize profits \emph{during} the learning phase. Finally, there are several variants of our model to be considered in future work, such as multiple buyers with different utility functions, and stochastic budgets that vary daily.

\newpage
\clearpage

\bibliographystyle{plain}
\bibliography{bib}


\end{document}

%% file: introduction.tex
We consider algorithmic and learning-theoretic aspects of the classical {\em revealed preferences problem\/}.
In this problem, a consumer has a fixed but unknown
utility function $u$ over $n$ goods, and is price sensitive. At each period $t$,
she observes a price vector $p^t$ and purchases a (possibly fractional) bundle of goods $x^t$
to maximize her utility given her budget (i.e. $x^t \in \arg\max_{x \cdot p^t \leq B} u(x)$).
Given a sequence of $T$ observations $(p^1,x^1),\ldots, (p^T,x^T)$ the revealed preferences problem (introduced by \cite{sam}; see \cite{varian} for a survey)
is to determine whether the observations are \emph{consistent} with a consumer optimizing any utility
function $u$, subject to some mild constraints (e.g. monotonicity of $u$). In this paper, however, we have
different and stronger objectives, motivated by what power the merchant has to set prices. We consider two scenarios:

\textbf{\bf\em \emph{(}Price-Setting\emph{)}} First, we consider a monopolist merchant who has the power to set prices as he wishes (without fear of losing the consumer to competition).
In this setting, we consider the natural goal of merchant profit maximization. The merchant has a fixed unit \emph{cost} associated with each good,
and his profit, when the consumer buys a bundle $x$, is his revenue minus the cost of the bundle purchased.
The merchant wishes to adaptively set prices so as to minimize his costs, which in turn maximizes his profits. Every round, the consumer purchases her utility
maximizing bundle subject to the merchant's prices and her budget constraint. If the merchant \emph{knew} the consumer's utility function,
he could optimally set prices, but he does not --- instead, the merchant faces a {\em learning\/} problem.
For the case when the consumer has a linear utility function, we give an efficient algorithm for
setting prices to quickly learn the consumer's utility function and then exploit this knowledge to set profit-maximizing prices.

\textbf{\em \emph{(}Exogenous Prices\emph{)}} Second, we consider a merchant who cannot unilaterally set prices, but instead must react to 
a stream of exogenously chosen prices.
This setting is relevant to a seller of commodity goods, or the owner of a franchise that must set prices given by the parent company.
Despite his lack of control over prices, this merchant would nevertheless like to be able to \emph{predict} which bundle the consumer is going to buy in the next period
(e.g. to optimize inventory or streamline the supply chain). The problem remains that the consumer's utility function is unknown,
and now the price sequence is also unknown and arbitrary (i.e. it could in the worst case be chosen adaptively, by an adversary).
In this setting, when the consumer has a linear utility function, we give an efficient algorithm with a small \emph{mistake bound} ---
in other words, even against an adaptively chosen set of price vectors, in the worst case over consumer utility functions,
our algorithm makes only a bounded number of mistaken predictions of bundles purchased.

We note that there are a variety of scenarios that fall under the two frameworks above. These include sponsored search or
contextual advertising on the web (where advertisers typically must obey periodic budget constraints, and prices are set exogenously 
by the bids of competitors or endogenously by an ad exchange or publisher); consumers who regularly receive gift certificates which can only be
used for purchases from a single
merchant such as Amazon, who in turn has price-setting powers;
consumers with poor self-control who repeatedly spend all of their disposable income in
a single category of goods (e.g. alcohol) from a price-setting merchant (e.g. local liquor store);
and crowdsourcing or labor management settings where a manager (merchant) can set rewards or payments for a set of daily
tasks, and workers (consumers) with a budget of time or effort select tasks according to the incentives and their own abilities.

From a learning-theoretic point of view, the price-setting framework can be viewed as a form of {\em query\/} model,
since the merchant is free to experiment with prices (both for the purpose of learning about the consumer, and subsequently in order
to optimize profits); while the exogenous price model falls into the large literature on adversarial, worst-case online learning.
The learning problem we consider in both settings is unusual, in that the ``target function'' we are trying to
learn is the vector-valued $\arg\max$
(optimal bundle) of the consumer's utility function. Additionally, despite the linearity of the consumer's utility function,  the merchant's 
reward function is a non-convex function of prices, which further complicates learning. 

Our major assumptions --- that consumers always spend their entire budget, that there is one divisible unit of each good available in each round, and that consumers repeatedly return to the same merchant --- are standard in the classic revealed preferences model (see e.g. textbook treatments in \cite{mwg} and \cite{Rub12}). In order to provide efficient learning algorithms in this setting, we necessarily impose additional restrictions on the form of the consumer's utility function.  In particular, we assume that the utility function is linear, and that the coefficients are discretized and lower-bounded. 
The discretization assumption is necessary to learn the utility function \emph{exactly}, which is required for the merchant's optimization. Even if two functions differ by an arbitrarily small amount, they can induce the consumer to buy very different bundles. We also assume an
upper bound on prices, and without loss of generality we rescale this
upper bound to be 1. Without such an assumption, the merchant could
maximize his profits by setting all prices to infinity. Unbounded
prices are neither founded in reality, nor lead to an
interesting optimization problem. 

\subsection{Our Results}

We first consider the case of a monopolist merchant who has the ability to set prices arbitrarily, and is facing a consumer with an
unknown linear utility function. In this setting, we give an algorithm with bounded regret with respect to the optimal (profit-maximizing) set of prices in hindsight.
Our argument proceeds in two steps. We first show that, \emph{if we knew} the consumer's utility function $u$, then we could efficiently
compute the optimal profit-maximizing prices $p^*$:

\begin{theorem1}[Informal]
There is an efficient algorithm \emph{(}running in time $O(n^2\log n)$\emph{)}, which given as input the linear consumer utility function $u$, outputs the profit-maximizing prices $p^*$.
\end{theorem1}

The analysis of this algorithm first assumes we know only the {\em set\/} of goods purchased by the consumer under
optimal prices (but not the optimal prices themselves), and
introduces a family of linear programs with one free parameter. We then show there is a small set of
values for this parameter, one of which yields the optimal prices. 

Note that although the \emph{consumer}'s optimization problem when selecting a bundle to purchase given prices
is simply a fractional knapsack problem, the problem of computing optimal prices is substantially more complex.
The optimal price vector $p^*$ is actually a subgame perfect Nash equilibrium strategy for the merchant in a two-stage extensive form
game between the merchant and the consumer (the merchant first picks prices, and then the consumer best responds).
Viewed in this way, the fractional knapsack problem that the consumer solves at the second stage is simply her best response function; what
we give is an algorithm for computing the merchant's subgame perfect equilibrium strategy in the first stage of this game. Note that doing this in polynomial time is non-trivial, because the merchant's strategy space is continuous (and even after discretization, is exponentially large).

We next give an algorithm that learns the consumer's unknown linear utility function by making price queries.
A price query specifies a price vector, and receives in return the bundle purchased by the consumer.
The key metric of interest here is the number of queries necessary to learn the utility function. We prove the following theorem:

\begin{theorem2}[Informal]
There is an efficient algorithm that learns, after at most $O(n)$ price queries, the utility coefficients for all goods except those that are so 
preferred they will be bought regardless of prices.
\end{theorem2}

This algorithm has two phases. In the first phase, after setting all prices to 1 (the maximum possible price),
the algorithm gradually lowers the price of unpurchased items in succession until they are purchased,
thus learning the ratio of their utility coefficient to that of the least preferred good purchased under the price vector of all 1s. Note that learning a \emph{relative} rather than exact coefficient value is sufficient because the bundle bought by the consumer (i.e. the solution to the knapsack problem) is unchanged by a scaling of the coefficients. Indeed, because of this property, it is impossible for any algorithm to learn the exact values. 

The harder coefficients to learn are those corresponding to goods purchased even when all prices are 1 --- these
are the consumer's most preferred goods. Some of these are learned by gradually lowering the prices of unpurchased goods
until a switch of purchased goods occurs;
for the ones that cannot be learned via this procedure, we prove that they are so favored they will be purchased
under any prices, so learning their coefficients is not necessary for price optimization.

Finally, we put these two algorithms together to prove our first main result:

\begin{theorem3}[Informal]
There is a price-setting algorithm that, when interacting with a consumer with an unknown linear utility
function for $T$ rounds, achieves regret $O(n^2/T)$ to the profit obtained by the optimal \emph{(}profit-maximizing\emph{)} price vector.
\end{theorem3}

In the last part of the paper, we consider the case of a commodity merchant who does not have the power to set prices.
The merchant wishes to predict the bundles that a consumer with an unknown linear utility function will buy, in the face of a stream of
arbitrary price vectors.
Here, the main quantity of interest is how many mistakes we make (by predicting the incorrect bundle)
in the worst case over both consumer utility functions and sequences of price vectors. We call this the {\em mistake
bound\/} of the algorithm (by analogy to the mistake bounded model of learning). Here we prove our second main result:

\begin{theorem4}[Informal]
There exists a polynomial time algorithm in the online exogenous price model that 
has a mistake bound of $O(n^2)$ with high probability.
\end{theorem4}

\subsection{Related Work}

The work most directly related to our results is the recent paper of Balcan et al. \cite{BDMUV14},
which was conducted independently and concurrently.
They study the problem of learning from
revealed preferences in various settings, including in a query model
related to the model we study here. Our ``price queries'' differ
slightly from the queries in the Balcan et al. model, in that our
learner can only specify prices, whereas the learner from Balcan et
al. can specify prices, as well as the consumer's budget with each
query. However, the main distinction between our work and theirs is
that our goal is profit maximization (even if we do not exactly learn
the buyer's utility function), and the goal of Balcan et al \cite{BDMUV14} is to exactly
learn the buyer's utility function --- they do not consider the
profit maximization problem.

More broadly, there is a long line of work on the revealed preference problem, which was first introduced by Samuelson \cite{sam}. For a textbook introduction to the standard model, see \cite{mwg} and \cite{Rub12}, and for a survey of recent work, see \cite{varian}. Most previous efforts have focused on the construction of utility functions that explain a finite sequence of price/bundle observations. Afriat's Theorem \cite{afriat} is the seminal result in this field, and proves that a sequence of observations is \emph{rationalizable} (i.e. can be explained by a utility function) if and only if the sequence is rationalizable by a piecewise linear, monotone, concave utility function. The proof is by construction, and therefore yields a ``learning'' algorithm for this large class of utility functions. However, the hypothesis generated has description length proportional to the number of observations, and hence although it can \emph{explain} previous observations, it usually does not generalize to \emph{predict} the bundles purchased given new price vectors.

The problem of finding a utility function that is both consistent and predictive was first considered by Beigman and Vohra \cite{vohra}, who formalize the statement that ``Afriat's Theorem Learners'' do not generalize. \cite{vohra} studies a PAC-like learning model in which they assume a distribution over observations and seek to find a hypothesis that performs well on future observations drawn from the same distribution. Their results essentially show that it is only possible to find predictive hypotheses if we restrict the class of allowable utility functions beyond those that are rationalizable. Roth and Zadimoghaddam \cite{roth} extended this line of work by providing computationally efficient learning algorithms for two specific classes of utility functions --- linear and linearly separable and concave utility functions.  Cummings, Echenique, and Wierman \cite{CEW14} consider the revealed preferences problem when the consumer can strategically choose bundles to subvert the merchant's learning.  In this setting, they show that without assuming the consumer's utility function is linearly separable, the merchant is unable to learn anything.  Like this previous work, we also seek to find predictive hypotheses for the class of linear utility functions, but we consider two new learning models: one in which prices are directly controlled, rather than observed (which corresponds to a query model of learning), and furthermore we wish to learn optimal prices;
and one in which prices are chosen adversarially and adaptively, and arrive online (which corresponds to online learning in the mistake bound model).

Our results in the second model are inspired by the classic halving algorithm for the online learning setting, which is credited to Littlestone \cite{littlestone}. The approach is simple but powerful: given a finite set of hypotheses and a sequence of observations, predict according to the majority of remaining hypotheses, and then discard all hypotheses that made a mistake. To implement the algorithm efficiently, we instead maintain a continuous hypothesis space from which we predict using a randomly sampled hypothesis (rather than predicting using a majority vote). We track the volume of the hypothesis space (rather than the number of consistent hypotheses), and show that after a bounded number of mistakes, we must have learned one coefficient of the consumer valuation function. 


%% file: preliminaries.tex
We consider a set of $n$ divisible goods that a merchant wishes to sell to a consumer. We represent a \emph{bundle} of goods $x \in [0,1]^n$ by a vector specifying what fraction of each of the $n$ goods is purchased. The consumer has an unknown utility function $u:[0,1]^n\rightarrow \mathbb{R}$ specifying her utility for each possible bundle. The merchant has the power to set prices (one for each good), also represented by a vector $p \in [0,1]^n$ (we normalize so that the price of every good $i$ is $p_i \leq 1$). Written in this way, the \emph{price of a bundle} $x$ is simply $x\cdot p= \sum_{i=1}^n p_i\cdot x_i$. Finally, the consumer behaves as follows: facing a price vector $p$, the consumer purchases her most preferred bundle subject to a budget constraint $B \geq 0$. That is, she purchases a bundle in $\arg \max_{x \cdot p \leq B} u(x)$. There may not always be a unique utility-maximizing bundle for the consumer, and so we let $X(u,p,B) = \arg\max_{x \cdot p \leq B} u(x)$, the set-valued collection of utility-maximizing bundles. If $X(u,p,B)$ is a singleton set, we say that the consumer's choice is \emph{uniquely specified} by $p$. We assume the budget is fixed and known to the merchant (although if the budget were unknown, the merchant could learn it from a single price query).

We restrict our attention to linear utility functions, which are defined by a valuation vector $v \in \mathbb{R}^n$ such that $u(x) = x \cdot v$. We assume the valuations are discretized to some increment $\delta$; i.e. each $v_i \in \{0, \delta, 2\delta,\ldots,1\}$. For this family of utility functions, the consumer's optimization problem to compute $X(u, p, B)$ is a fractional knapsack problem. The capacity of the knapsack is $B$, and the weight and value of a good $i$ are $p_i$ and $v_i$, respectively. This problem can be solved greedily by ranking the goods in decreasing order of their $v_i/p_i$ (i.e. bang per buck) ratios, and then buying in this order until the budget is exhausted. Note that in the optimal bundle, there will be at most one fractionally purchased good. Since this ratio is important in many of our algorithms, given $u$ and $p$, we will denote $v_i/p_i$ by $r_i(u,p)$, or by $r_i$ when $u$ and $p$ are clear from context. If $r_i \leq r_j$ we say that the consumer \emph{prefers item $i$ to item $j$}. 

We consider two problem variants. In the first, the merchant has the power to set prices, and has a production cost $c_i \leq 1$ for each good $i$. Hence, the merchant's \emph{profit} when the consumer buys a bundle $x$ at prices $p$ is $x\cdot (p-c)$. It always improves the consumer's utility to saturate her budget and so $x \cdot p = B$ for any $x \in X(u,p,B)$ and $x \cdot (p - c) = B - x \cdot c$. Hence, maximizing the merchant's profit is equivalent to minimizing his costs $x \cdot c$. The merchant's goal is to obtain profit close to the maximum possible profit $\mathrm{OPT} = \max_{p \in [0,1]^n} \max_{x \in X(u,p,B)} x\cdot(p-c) = \max_{p \in [0,1]^n} \max_{x \in X(u,p,B)} B - x \cdot c.$

Note that solving this problem requires both learning something about the unknown utility function $u$, as well as the ability to solve the optimization problem given $u$. At every round $t$, the algorithm chooses some price vector $p^t$, and the consumer responds by selecting any consistent $x^t \in X(u,p^t,B)$. We measure our success over $T$ time steps with respect to our regret to the optimal profit the merchant could have obtained had he priced optimally at every round, which is defined as $$\mathrm{Regret}(p^1,\ldots,p^T) = \text{OPT} - \frac{1}{T}\sum_{t=1}^T x^t \cdot(p^t-c)$$

In the second variant, we view price vectors $p^1,\ldots,p^T$ arriving one at a time, chosen (possibly adversarially) by Nature. In this setting, the merchant has no control over the bundle purchased by the consumer, and wish only to predict it. At each time step $t$, after learning $p^t$, we get to predict a bundle $\hat{x}^t$. Following our prediction, we observe the bundle $x^t \in X(u,p^t,B)$ actually purchased. We say that we make a \emph{mistake} if $\hat{x}^t \neq x^t$, and our goal is to design an algorithm that makes a bounded number of mistakes in the worst case over both $u$ and the sequence of prices $p^1,\ldots,p^T$. 
 



%% file: optprices.tex
\newcommand{\bx}{x}
\newcommand{\bp}{p}
\newcommand{\bc}{c}

In this section we assume that all the coefficients $v_i$ of the consumer's utility function are known to the merchant. Even then, it is not clear \emph{a priori} that there exists an efficient algorithm for computing a profit-maximizing price vector $\bp$. As previously mentioned, the optimal prices are a subgame perfect Nash equilibrium strategy for the merchant in a two-stage extensive form game, in which the merchant has exponentially many strategies. Straightforwardly computing this equilibrium strategy via backwards induction would therefore be inefficient. We show that nevertheless, this task can be accomplished in time only (nearly) quadratic in the number of goods. 

The key to the algorithm's efficiency will stem from the observation that there exists a restricted family of pricing vectors $\mathcal{P} \subset [0,1]^n$ containing a (nearly) profit-maximizing vector $\bp^*$. This subset $\mathcal{P}$ will still be exponentially large in $n$, but will be ``derived'' (in a manner which will be made more precise) from a small set of vectors $\bp^{(1)},...,\bp^{(n)}$. This derivation will allow the algorithm to efficiently search for $\bp^*$. We define $\bp^{(k)}$ by letting $\bp^{(k)}_i = \min(v_i/v_k, 1)$. In other words, the price of every good whose value is less than the $k$th good is set to the ratio $v_i/v_k$. Otherwise, if $v_i > v_k$, the price of good $i$ in $\bp^{(k)}$ is set to the ceiling of $1$.    

To understand the operation of the algorithm, consider the consumer's behavior under the prices $\bp^{(k)}$. Any good priced at $v_i/v_k$ will have a bang per buck ratio $r_i = v_k$. Therefore, the consumer's choice is not uniquely specified by $\bp^{(k)}$ in general (since the consumer is indifferent between any of the previously mentioned goods). Moreover, the consumer's choice will have great impact on the merchant's profit since the goods between which the consumer is indifferent might have very different production costs $\bc_i$. The algorithm therefore proceeds by computing, for each $\bp^{(k)}$, which bundle $\bx^{(k)}$ the merchant \emph{would like} the consumer to purchase under $\bp^{(k)}$. More precisely, for each $k$, the algorithm computes $\bx^{(k)} \in \arg\max_{\bx \in X(u,\bp^{(k)},B)} \bx \cdot (\bp^{(k)} - \bc)$. Note that if the merchant were to actually play the price vector $\bp^{(k)}$, the consumer would be under no obligation in our model to respond by selecting $\bx^{(k)}$. Therefore, the final step of the algorithm is to output a price vector which attains nearly optimal profit, but for which the consumer's behavior is uniquely specified. 

The pseudocode is given in Algorithm \texttt{OptPrice}.
The analysis proceeds by proof of three key facts. (1) For some $k$, the optimal profit is attained by $(\bp^{(k)},\bx^{(k)})$, or rather, $\mathrm{OPT} = \bx^{(k)} \cdot (\bp^{(k)} - \bc)$ for some $k$. (2) Given any $\bp^{(k)}$, $\bx^{(k)}$ can be computed efficiently (in $O(n \log n)$ time). Finally, (3) there is some price $\hat{\bp}$ for which the consumer's choice $\bx$ is uniquely specified, and where $\bx \cdot (\hat{\bp} - \bc)$ is close to $\mathrm{OPT}$.  

\begin{varalgorithm}{OptPrice($B,v,c,\epsilon, \delta$)}
\caption{}
\label{alg:optprices}
\begin{algorithmic}
\For {$k = 1$ to $n$}
\State $p^{(k)}_i = \min(v_i/v_k, 1)$ for all $i$
\State $x^{(k)} = \arg\max_{x \in X(u,p^{(k)},B)} x \cdot (p^{(k)} - c)$
\Comment{$O(n \log n)$ computation}
\State $\mathrm{Profit}(k) = x^{(k)} \cdot (p^{(k)} - c)$
\EndFor
\State $k_{\max} = \arg\max_k \mathrm{Profit}(k)$
\State $p^* = p^{(k_{\max})}$, $x^* = x^{(k_{\max})}$
\For {$i = 1$ to $n$}
  \If {$x^*_i = 0$}
    \State $\hat{p}_i = 1$
  \ElsIf{$x^*_i = 1$}
    \State $\hat{p}_i = p^*_i - \epsilon$
  \Else 
    \State $\hat{p}_i = p^*_i - \delta\epsilon$
  \EndIf
\EndFor
\State \textbf{return} $\hat{p}$
\end{algorithmic}
\end{varalgorithm}

\begin{theorem1}
\label{thm:offline}
Algorithm \emph{\texttt{OptPrice}} \emph{(}which runs in time $O(n^2\log n)$\emph{)}, takes coefficients $v_1,\ldots,v_n$ as input and for any $\epsilon > 0$, computes
prices $\hat{p}$ for which the consumer's choice $\hat{x}$ is \emph{uniquely specified} and achieves profit $x (\hat{p} - c) \geq \emph{OPT} - \epsilon$.
\end{theorem1}

\proof
We prove the above theorem by establishing the three key facts listed above. The first key lemma establishes that optimal profit is attained by some $(\bp^{(k)}, \bx^{(k)})$ for some $k$. 

\begin{lemma}
\label{lem:price}
Let $\bp^{(k)}$ be the pricing vector such that $\bp_i^{(k)} = \min(v_i/v_k, 1)$. For any consumer utility parameterized by $(u,B)$, there exists some $k$ and an $\bx \in X(u,\bp^{(k)},B)$, such that $\mathrm{OPT} = \bx \cdot (\bp^{(k)} - \bc)$. 
\end{lemma}

\proof
Fix $u$ and $B$. Consider a profit-maximizing price $\bp^*$, and corresponding bundle $\bx^* \in X(u,\bp^*, B)$, so that $\mathrm{OPT} = \bx^* \cdot (\bp^* - \bc)$. Let $O = \{i : \bx^*_i > 0\}$ be the set of purchased goods in $\bx^*$. If there is a fractionally purchased good in $\bx^*$, we denote its index by $f$ (i.e. $0 < \bx_f^* < 1$). 

We note that there must exist a threshold $\tau$ such that $r_i(u,\bp^*) \leq \tau$ whenever $i\not\in O$ and $r_i(u,\bp^*) \geq \tau$ whenever $i \in O$. In words, in order for the bundle $\bx^*$ to maximize the consumer's utility, the bang per buck of every purchased good must be at least as large as the bang per buck of every unpurchased good. If there is a fractional good in $\bx^*$, we take $\tau$ to be $r_f(u,\bp^*)$. Otherwise, we take it to be $\max_{i \not\in O} r_i(u, \bp^*)$, the bang per buck of the most desirable unpurchased good. 

Given $\tau$, we can write the linear program in \eqref{eq:lp}. We claim that any solution to this LP is also a profit-maximizing price. More precisely, if $\bp^{(LP)}$ is a solution to the LP, then there is an $\bx^{(LP)} \in X(u,\bp^{(LP)},B)$ such that $\mathrm{OPT} = \bx^{(LP)}(\bp^{(LP)}-c)$. The following argument proves this claim.

\begin{equation}
\label{eq:lp}
\begin{aligned}
& \max && \sum_{i \in O} \bp_i && \\
& \; \text{s.t.} && v_i/\bp_i \geq \tau && \forall i \in O \\
&&& v_i/\bp_i \leq \tau &&  \forall i \notin O \\
&&& \bp_i \leq 1 && \forall i \in [n]
\end{aligned}
\end{equation}

We can straightforwardly characterize any solution $\bp^{(LP)}$ to the LP. Note that the constraints on $\bp_i$ are disjoint, and therefore for each $i \in O$, $\bp_i$ can increased independently until a constraint is saturated. Thus, the LP is optimized by setting $\bp_i^{(LP)} = \min\{v_i/\tau, 1\}$ for each $i \in O$, and $\bp_i^{(LP)} \geq v_i/\tau$ for each $i \not\in O$ (which is always possible since $\bp^*$ is a feasible solution). 

The LP constraints imply that, under $\bp^{(LP)}$, the consumer (weakly) prefers any good in $O$ to any good not in $O$. Moreover, if there is a fractional good, the definition of $\tau$ ensures that any good in $O \setminus \{f\}$ is preferred to $f$. Finally, we know that $\sum_{i \in O} \bp_i^* \leq \sum_{i \in O} \bp^{(LP)}_i$ since $\bp^*$ is a feasible solution to the LP. This, along with the preference ordering on goods imposed by $p^*$, tells us that the consumer can saturate her budget at least as quickly under $\bp^{(LP)}$ as under $\bp^*$. In other words, under $\bp^{(LP)}$, the consumer may saturate her budget \emph{before} purchasing all goods in $O \setminus \{f\}$, or may require a smaller allocation of good $f$, but she will never require a larger allocation of good $f$ or purchase any goods not in $O$.  More precisely, there exists an $\bx^{(LP)} \in X(u,\bp^{(LP)},B)$ such that point-wise $\bx^{(LP)}_i \leq \bx^*_i$. Thus, $\mathrm{OPT} = \bx^* (\bp^* - \bc) = B - \bx^* \cdot \bc \leq B - \bx^{(LP)} \cdot \bc = \bx^{(LP)} \cdot (\bp^{(LP)} - \bc)$, and so $\bp^{(LP)}$ must be a profit-maximizing price. 
   
We have established that $\bp^{(LP)}$, which sets $\bp_i = \min(v_i/\tau,1)$, is profit-maximizing, and now need only to show that we can take $\tau = v_k$ for some $k$.  Consider again a profit-maximizing price $\bp^*$ and corresponding bundle $\bx^*$. Notice that for any $i \not\in O$, we can modify prices to set $\bp_i^* = 1$, and $(\bx^*,\bp^*)$ will still maximize profit because this change only makes the unpurchased goods more undesirable. Notice that with this modification, $\max_{i \not\in O} r_i(u,\bp^*) = \max_{i \not\in O} v_i = v_{k^*}$ for some $k^*$. Next consider modifying the price of the fractional good. The price of $f$ can be increased to $v_f/v_{k^*}$ if $v_f/v_{k^*} < 1$, and $1$ otherwise. This increases the price of $f$ while still keeping it as the fractionally purchased good, thus reducing the merchant's cost. This will result in either $r_f = v_{k^*}$ or $r_f = v_f$. In either case, there exists a $(\bx^*,\bp^*)$ such that $\tau = v_k$ for some $k$. Thus an optimal price $\bp^{(LP)}$ can be derived by searching over at most $n$ possible values of $\tau$, and sets $\bp_i = \min(v_i/v_k, 1)$ for some $v_k$.  
\qed(Lemma 1) \\

Lemma \ref{lem:price} establishes that for some $k$, $\bp^{(k)}$ is optimal, in the sense that there exists an $\bx^{(k)} \in X(u,\bp^{(k)},B)$ such that $\mathrm{OPT} = \bx^{(k)}(\bp^{(k)} - c)$. The algorithm proceeds by computing for each such $\bp^{(k)}$, the most profitable bundle (for the merchant) that the consumer (who is indifferent between all bundles in $X(u,\bp^{(k)},B)$) could purchase. $X(u,\bp^{(k)},B)$ is potentially a very large set. For example, if the consumer has identical values for all goods (i.e. $v_i \equiv c$ for all $i$), and $\bp^{(k)}$ is the all-1s vector, then $X(u,\bp^{(k)},B)$ contains any budget-saturating allocation. Despite the potential size of this set, Lemma 2 shows that computing $\max_{\bx \in X(u,\bp^{(k)},B)} \bx \cdot (\bp - \bc)$ simply requires solving a fractional knapsack instance, this time from the merchant's perspective. 

\begin{lemma}
For any $\bp,u,$ and $B$, $\max_{\bx \in X(u,\bp,B)} \bx \cdot (\bp - \bc) = B - \bx \cdot \bc$ can be computed in $O(n \log n)$ time.
\end{lemma} 
\proof
Let $\bp$ be an arbitrary price vector. The bundle $\max_{\bx \in X(u,\bp,B)} B - \bx \cdot \bc$ can be computed as follows. First sort the $r_i$ in decreasing order, so that $r_{i_1} \geq ... \geq r_{i_n}$. The consumer will buy goods in this order until the budget $B$ is exhausted. Thus, we can simulate the consumer's behavior by iteratively ``buying'' goods and decrementing the budget. The consumer's behavior is uniquely specified unless we reach some sequence of goods with $r_{i_j} = r_{i_{j+1}} = \ldots = r_{i_{j+d}}$, with $B'$ budget remaining, where $\sum_{l=0}^{d} \bp_{i_{j+l}} > B'$. In other words, the consumer's behavior is unique unless she is indifferent between some set of goods, and can select different subsets of these goods to exhaust her remaining budget $B'$.

In that case, we know that for any bundle $x \in X(u,\bp,B)$, $\bx_{i_l} = 1$ if $l < j$, and $\bx_{i_l} = 0$ if $l > j+d$. For the remaining goods, the merchant's profit is maximized when $\bx \cdot \bc$ is minimized. This occurs when the consumer saturates the remaining budget $B'$ while minimizing the cost $\bc$ to the merchant. This is an instance of min-cost fractional knapsack wherein the size of the goods are $\bp_{i_j},...,\bp_{i_{j+d}}$ and the cost of the goods are $\bc_{i_j},...,\bc_{i_{j+d}}$. A solution to this problem can be computed greedily, so the most profitable bundle for $\bp$ can be computed with at most two sorts (first for $r_i$ then for $\bp_i/c_i$). 
\qed (Lemma 2)\\

Finally, to induce the consumer to buy the bundle $x^*$, rather than another member of $X(u, p^*, B)$, we perturb the vector $p^*$ slightly, and show that this has an arbitrarily small effect on profit.

\begin{lemma}
For any $\epsilon > 0$, there exists a price vector $\hat{p}$ that uniquely specifies a bundle $\hat{x}$ that satisfies $\hat{x} \cdot (\hat{p} - c) \geq \emph{OPT} - \epsilon$.
\end{lemma}
\proof
Recall that the merchant would like the consumer to purchase the bundle $x^*$, which is a member of the set $X(u, p^*, B)$. Even if the merchant sets prices at $p^*$, there is no guarantee that the consumer will purchase $x^*$ rather than some other bundle in the set. Our goal is to output a vector $\hat{p}$ that results from perturbing $p^*$ slightly, so that the consumer will always purchase some $\hat{x}$ that is arbitrarily close to $x^*$. For any good $i$ such that $x^*_i = 0$ (a good that consumer should not buy at all), we simply set $\hat{p}_i = 1$. For any good $i$ such that $x^*_i = 1$ (a good that that the consumer should buy in its entirety), we set $\hat{p}_i = p^*_i - \epsilon_0$. Finally, for any good $i$ such that $0 < x^*_i < 1$ (a good that the consumer should buy fractionally), we set $\hat{p}_i = p_i^* - \delta\epsilon_0$. 

We claim that these perturbations ensure that the consumer will buy goods in the order desired by the merchant. First, note that these prices will never cause an item that was not purchased in $x^*$ to be preferred to anything that was purchased in $x^*$.  Consider an item $i$ such that $x^*_i = 0$.  Before we perturbed prices, all purchased goods were (weakly) preferred to $i$.  Now these purchased items have even lower prices, making them (strictly) more appealing, while $i$ retains its same bang per buck. Thus the purchased goods in $x^*$ are all now (strictly) preferred to $i$. 

Second, the new prices will never cause an item that was fully purchased in $x^*$ to be purchased after the fractional good of $x^*$. Consider a good $i$ such that $x^*_i = 1$, and the fractional good $f$ such that $0 < x^*_f < 1$. Suppose we perturb prices such that $\hat{p}_i = p^*_i - \epsilon$, and $\hat{p}_f = p^*_f - \epsilon/\gamma$, and therefore must decide how large $\gamma$ needs to be to satisfy our claim. For our claim to be true, we need the new bang per buck of good $i$ is strictly higher than the new bang per buck of good $f$.  That is, we want:
$$\frac{v_i}{p^*_i - \epsilon} > \frac{v_f}{p^*_f - \epsilon/\gamma}$$
which is equivalent to:
$$ \gamma > \frac{v_i \epsilon}{p^*_f v_i - v_f(p^*_i - \epsilon)}$$
Recall that $i$ was purchased before $f$ under $p^*$, so $i$ previously had a bigger bang per buck than $f$, and therefore $p^*_f v_i - v_f p^*_1 \geq 0$. So it is sufficient to choose
$\gamma > \frac{v_i \epsilon}{v_f \epsilon} = \frac{v_i}{v_f}$. We know that each value lies in the range $[\delta, 1]$, so this fraction is maximized at $v_i = 1, v_f = \delta$, and we can take $\gamma = 1/\delta$.

Next we must show that the new prices only cause a small reduction in profit. We have decreased each price by at most $\epsilon_0$, so the consumer might have up to an additional $n\epsilon_0$ budget to spend. Recall that optimal prices are chosen by the algorithm to be $p_i = \min(v_i/v_k, 1)$ for each $i$ and for some $k$. Because values are discretized and lower-bounded, the minimum price possible is therefore $\delta$. Consider setting $\epsilon_0 = \delta \epsilon/n$, which yields at most $\delta \epsilon$ additional budget. Then the consumer can afford to purchase at most an additional $\delta \epsilon/\delta = \epsilon$ fraction of a good. In the worst case, if this good is of maximum cost 1, the merchant will incur an additional cost of $\epsilon$. 
\qed (Lemma 3) \\
\qed (Theorem 1)

%% file: learnval.tex
We now provide a query algorithm for learning the coefficients $v_i$ of the consumer's utility function. 
For the analysis only, we assume without loss of generality that the goods are ordered by decreasing value, i.e. $v_1 > \ldots > v_n$.  
Our  algorithm can learn the values in some suffix of this unknown ordering;
the values that cannot be learned are irrelevant for setting optimal prices, since those goods will always be purchased by the consumer.

\begin{theorem2}
\label{thm:learnval}
Algorithm \emph{\texttt{LearnVal}}, given the ability to set prices, after at most $O(n \log((1-\delta)/\delta^2))$ price queries, outputs the ratio $v_i/v_n$ for all goods $i$ except those that will be bought under all price vectors, when values are discretized by $\delta$. 
\end{theorem2}

\begin{varalgorithm}{LearnVal($\delta$)}
\caption{}
\label{alg:learnval}
\begin{algorithmic}
\State $p_i = 1 \quad \forall i$
\Comment{First price query}
\State $x \leftarrow \texttt{consumer}^1(p)$
\While {$\neg \exists i \text{ such that } 0 < x_i < 1$}
\Comment{Find some fractionally bought good}
\State Choose $k$ such that $x_k = 0$
\State $p_k = p_k - \delta$
\State $x \leftarrow \texttt{consumer}(p)$
\EndWhile
\State $j = i$ \Comment {$j$ is least-preferred purchased good}
\For {$k = j+1$ to $n$} \Comment{Learn $s_k$ for $j+1, \ldots, n$}
\While {$x_k = 0$}
\State $p_k = p_k - \delta$
\State $x \leftarrow \texttt{consumer}(p)$
\EndWhile
\State $v_k/v_j = p_k$
\EndFor
\State $s_k = (v_k/v_j)/(v_n/v_j) \quad \forall k \geq j$
\Comment {Renormalize ratios}
\For {$k = j-1$ to $1$} \Comment{Learn $s_k$ for $j-1, \ldots, 1$}
\For {$\alpha = 1$ to $\delta$ (in increments of $\delta$)}
\State $p_i = 1 \quad i \leq k$
\State $p_i = \alpha s_i \quad \forall i > k$
\State $x \leftarrow \texttt{consumer}(p)$
\If {$x_k > 0$}
\State $s_k = 1/\alpha$
\State \textbf{break}
\EndIf
\EndFor
\If {$s_k$ is undefined}
\Comment{$k$ was always bought}
\State \textbf{break}
\EndIf
\EndFor
\State \textbf{return} $s$
\State
\State $^{1}$The notation $x \leftarrow \texttt{consumer}(p)$ specifies the bundle bought by the consumer at prices $p$.
\end{algorithmic}
\end{varalgorithm}

\proof
Algorithm \texttt{LearnVal} proceeds by making price queries. That is, on each day we choose a particular price vector, observe the bundle purchased at those prices, and then use this information as part of the learning process. For our first query, we set $p_i=1$ for all $i$ and observe the bundle $x$ bought by the consumer. Let $j$ be the index of the least-preferred good that the consumer purchases under this price vector. If the consumer buys some good $i$ fractionally (which the algorithm can observe), then $j = i$. Otherwise, to learn $j$, we can incrementally lower the price of some good $k$ that the consumer did not purchase, until $k$ is purchased instead of another good $i$. Then we have learned $j = i$.

In the next phase of this algorithm, we will learn the ratio $v_k/v_j$ for all goods $k > j$ that were not originally purchased. To do so, we lower $p_k$ (while keeping all other prices at 1) until item $k$ is purchased.  This will occur when $v_k/p_k = v_{j}/p_{j}$, or $v_k/v_{j} = p_k$. Recall that we assume each $v_i$ is discretized to a multiple of $\delta$. 
Therefore to guarantee that we learn the ratio $v_k/v_j$ \emph{exactly}, we must learn the ratio up to a precision of $\min_{k \neq k'} |(v_k - v_{k'})|/v_j$. 
This quantity is minimized at $v_k = v_{k'} + \delta$ and $v_j = 1$ (because $v_j \leq 1$), so it is sufficient to learn the ratio to within $\pm \delta$. 
Thus if we perform a discrete binary search on $p_k$,
it will require $O(n \log(1/\delta))$ steps to exactly identify the desired ratio. 
Finally, we renormalize the ratios we have learned in terms of $v_n$. That is, for all $k \geq j$, we define $s_k = v_k/v_n = (v_k/v_{j}) / (v_n/v_j)$. 

We next attempt to learn $s_k = v_k/v_n$ for all $k < j$. These are the most preferred goods that were originally purchased under the all-1s price vector. 
We learn $s_k$ inductively in decreasing order of $k$, so as we learn $s_k$, we already know the value $s_i$ for all $i > k$. 
The goal is to now set prices so that the consumer will be indifferent to all goods $i > k$ (i.e. will have a tie in the bang per buck for all these goods).
The bang per buck on these goods is initially set low, and gradually raised by adjusting prices.  At some critical point, a switch in the behavior of the consumer will be observed in which $k$ is no longer purchased, and $s_k$ is learned.

\begin{figure}[h!]
\centering
\includegraphics[width=0.6\textwidth]{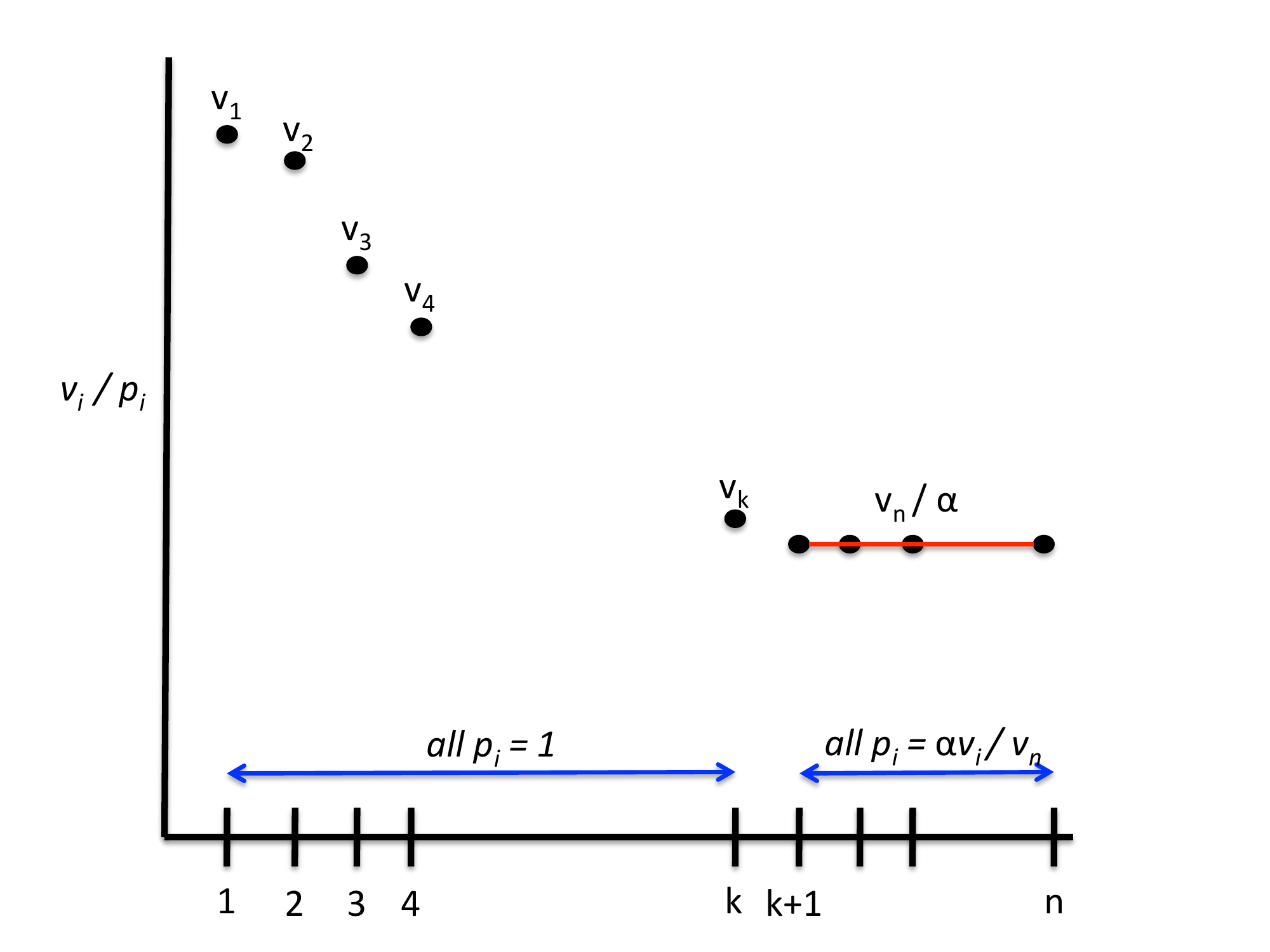}
\caption{\small An illustration of the process by which we search over $\alpha$. The figure depicts the situation just before hitting the critical point $\alpha^*(k)$. 
By lowering $\alpha$ slightly, we will raise the bang per buck $v_n/\alpha$ of all goods $k+1, \ldots, n$ (represented by the horizontal red line) 
to be slightly larger than $v_k$, 
and the consumer now prefers these goods to good $k$.}
\label{fig:waterlevel}
\end{figure}

We therefore introduce a parameter $\alpha$, which controls the bang per buck ratio of goods $k+1, \ldots, n$. Define $p(\alpha,k)$ to be the price vector where $p_i = 1$ for $i \leq k$, and $p_i = \alpha s_k$ for $i > k$. Let $r(\alpha,k)$ denote the corresponding bang per buck vector; i.e. $r(\alpha,k)_i = v_i/p_i$. 
It is easy to see that $r(\alpha,k) = (v_1, \ldots, v_k, v_n/\alpha, \ldots, v_n/\alpha)$. Thus, by lowering $\alpha$, we lower the prices and raise the desirability of goods $k+1, \ldots, n$. This process is illustrated in Figure \ref{fig:waterlevel}, whereby gradually lowering $\alpha$ from 1, we eventually reach a first point
where goods $k+1,...,n$ are preferred to good $k$. This switch occurs when the bang per buck ratio of goods $k+1, \ldots, n$ equals the bang per buck ratio of $k$, i.e. $v_n/\alpha = v_k$, or $\alpha = v_n/v_k$. Our goal will be to identify this value of $\alpha$, which we denote $\alpha^*(k)$. Once we know $\alpha^*(k)$, we will have learned $s_k$, since $s_k = v_k/v_n = 1/\alpha^*(k)$. 
 
We will know that we have found $\alpha^*(k)$ when we identify the highest value of $\alpha$ for which some good goes from being purchased to unpurchased. This good must be $k$, because as we decrease $\alpha$, we increase the desirability of goods $k+1, \ldots, n$, so none of these goods will go from being purchased to unpurchased. Of goods $1, \ldots, k$, good $k$ is the least preferred, so this will be the first good to become unpurchased.

To learn the value of $\alpha^*(k) = v_n/v_k$ exactly, we must have a precision of $\min_{k \neq k'}|v_n(|1/v_k - 1/v_{k'}|)$. 
This quantity is minimized at $v_n = \delta$ (because we assume a lower bound on values, and therefore $v_n \geq \delta$), 
$v_k = 1$, and $v_{k'} = 1-\delta$, 
and the corresponding value is $\delta^2/(1-\delta)$. Thus, we should search for the critical $\alpha^*(k)$ 
over the interval $[0,1]$ in increments of this size. In the algorithm's implementation, we can perform a binary rather than linear search over the range of $\alpha$, which therefore requires requires $O(n \log((1-\delta)/\delta^2))$ steps.
Once we have identified the value of $\alpha^*(k)$ to within $\delta^2/1-\delta$, we set $s_k = 1/\alpha^*(k)$.

In this manner, we can inductively find the next ratio $s_{k-1}$ by equalizing preferences for the later goods
and searching for the critical $\alpha^*(k-1)$. The only problem that might arise is that good $k$ is sufficiently
valued, and the budget sufficiently large, such that good $k$ is always purchased no matter how low $\alpha$ is set.
Lemma \ref{lem:purchased} shows that if this occurs for some $k$, then
no matter how we price goods $1, \ldots, k$, the consumer will always purchase these goods in full.
Thus for the purposes of setting prices optimally for the merchant, it is unnecessary to learn the values of these goods, since he can set the corresponding prices to the maximum of 1.

\begin{lemma}
\label{lem:purchased}
If goods $1, \ldots, k$ are purchased at price vector $p(\alpha',k)$, where $\alpha' = \alpha^*(k)-\delta^2/(1-\delta)$, then goods $1, \ldots, k$ will be purchased at any price vector.
\end{lemma}

\proof
Consider the corresponding bang per buck vector $r(\alpha',k)$. For $i \leq k$, $r(\alpha',k)_i = v_i \geq v_k$, and for $i < k$, $r(\alpha',k)_i > v_k$. Thus, good $k$ has the lowest 
bang per buck of all goods. Since good $k$ is still purchased, all other goods $i \neq k$ must be purchased as well. Consider raising the price of some good $j$ from its price under $p(\alpha',k)$, so that good $j$ becomes less desirable. It must be that $j > k$, because all other prices are already maximized at 1, so this does not change the fact that goods $1, \ldots, k$ will be purchased. Next consider lowering the price of some good $j$. This simply frees up more of the consumer's budget and allows the consumer more purchasing power, so all goods will remain purchased. 
\qed (Lemma~\ref{lem:purchased}) \\

The last phase of the algorithm is the bottleneck, and so the overall running time (i.e. number of queries made) is $O(n \log((1-\delta)/\delta^2))$, which is linear in the bit description length of values.
\qed (Theorem 2)


%% file: puttogether.tex
We now combine the previous two sections into a complete online, no-regret algorithm. The informal description of Algorithm \texttt{ProfitMax} is as follows. First we use Algorithm \texttt{LearnVal} to learn all possible $s_i$ ratios. For any good $i$ for which we did not learn $s_i$, we set $p_i$ at 1. Then we apply Algorithm \texttt{OptPrice} to the subset of remaining goods, using the $s_i$ ratios as input. The following main result shows that this approach achieves no-regret.

\begin{varalgorithm}{ProfitMax($B, c, \delta, \epsilon, T$)}
\caption{}
\label{alg:complete}
\begin{algorithmic}
\State $\text{Profit} = 0$
\State $s \leftarrow \texttt{LearnVal}(\delta)$
\Comment{Learn $s_i$ ratios}
\For{$i = 1$ to $n$}
\If{$s_i = 0$}
\State $p_i = 1$
\EndIf
\EndFor
\State $v = \{s_i \mid s_i \geq 0\}$
\State $p \leftarrow \texttt{OptPrice}(B, v, c, \epsilon)$
\Comment{Compute optimal prices}
\While {$t \leq T$}
\Comment{Set prices optimally}
\State $x^t \leftarrow \texttt{consumer}(p)$
\State $\text{Profit} = \text{Profit} + x^t \cdot (p - c)$
\EndWhile
\end{algorithmic}
\end{varalgorithm}

\begin{theorem3}
\label{thm:overall}
For any $\epsilon > 0$, after $T$ rounds, Algorithm \emph{\texttt{ProfitMax}} achieves per-round regret $O((n^2/T)\log((1-\delta)/\delta^2))$ to the profit obtained by the optimal price vector (where $\epsilon$ is the additive approximation to the optimal profit and $\delta$ is the discretization of values).
\end{theorem3}

\proof
First we show that this algorithm correctly composes Algorithms \texttt{LearnVal} and \texttt{OptPrice} and indeed generates an approximately optimal price vector $p$. 
\begin{lemma}
\label{lem:optprice}
An approximately optimal pricing for goods $\{1, \ldots, n\}$ is obtained by setting $p_i = 1$ for all goods for which Algorithm \emph{\texttt{LearnVal}} could not learn $s_i$, and then applying Algorithm \emph{\texttt{OptPrice}} to the $s_i$ ratios of the remaining goods.
\end{lemma}
\proof
According to Lemma \ref{lem:purchased}, for any good $i$ for which Algorithm \texttt{LearnVal} could not learn $s_i$, the consumer will always purchase $i$, regardless of $p_i$. Thus, our profit is maximized by setting $p_i = 1$. Furthermore, recall that the bundle bought by the consumer is invariant to scaling, and so the consumer will buy the same bundle regardless of whether we price according to the $s_i = v_i/v_n$ ratios or the actual $v_i$ values. Because our profit depends only on the bundle bought, it is therefore sufficient to use Algorithm \texttt{OptPrice} to price approximately optimally for the $s_i$ ratios.
\qed (Lemma~\ref{lem:optprice}) \\

As the above lemma shows, every time we price according to $p$, we receive approximately optimal profit. In particular, according to Theorem 1, our profit will be less than OPT by at most $\epsilon$, for any $\epsilon > 0$. Furthermore, Algorithm \texttt{LearnVal} uses at most $O(n \log((1-\delta)/\delta^2))$ price queries, so there are $O(n \log((1-\delta)/\delta^2))$ days on which we might incur maximum regret. On any given day, the maximum possible profit is $B$, while the minimum possible is $B-n$, yielding a maximum regret of $n$. Thus, our overall per-step regret is bounded by $O((n^2/T) \log((1-\delta)/\delta^2) + \epsilon)$. Setting $\epsilon = (n^2/T) \log((1-\delta)/\delta^2)$ yields the bound in the theorem statement.
\qed (Theorem 3)


%% file: exogenous.tex
We now consider our second model, in which an arbitrary and possibly adversarially selected price vector arrives every day, and the goal of our algorithm is to predict the bundle purchased by the consumer. Recall that the motivating scenario for such a setting is a merchant who is forced to set prices according to the market or the choices of a parent company. At each day $t$, the algorithm observes a price vector $p^t$ and makes a prediction $\hat{x}^t$. The algorithm then learns the bundle purchased, $x^t \in X(u, p^t, B)$, and is said to \emph{make a mistake} if $x^t \neq \hat{x}^t$. Our goal is to prove an upper bound on the total number of mistakes ever made by our algorithm, in the worst case over price vectors and utility functions. We call such an upper bound a \emph{mistake bound}.

We first informally describe the algorithm, given more precisely as Algorithm \texttt{ExogLearnVal}. The algorithm maintains the set of valuation vectors $v$ consistent with the observations seen thus far: initially this feasible set is simply $C_0 = [0,1]^n$. At every round $t$, the observed pair $(p^t,x^t)$ provides a set of linear constraints which we add to further constrain our feasible set $C_t$. In particular, we have the following lemma, first observed in \cite{roth}.

\begin{lemma}[\cite{roth}]
\label{lem:constraints}
For any pair of goods $i,j \in [n]$ where $x^t_i > x^t_j$, it must be that $v_i/p^t_i \geq v_j/p^t_j$.
\end{lemma}

This lemma follows immediately from the fact that the vector $x^t$ is the solution to a fractional knapsack problem, for which the optimal algorithm is to buy goods in decreasing order of $v_i/p_i^t$.
The set of all such constraints learned so far at time $t$ forms the feasible set $C_t$, which is a convex polytope. The idea of the algorithm is to sample a hypothesis valuation function $v^t$ uniformly at random from $C_t$ at each stage, and predict using the bundle that would be purchased if the buyer had valuation function $v^t$. The event that this hypothesis makes a mistake in its prediction is exactly the event that the hypothesis is eliminated when we update the consistent set to $C_{t+1}$, and so the probability of making a mistake is exactly equal to the fraction of volume eliminated from our consistent set, which allows us to charge our mistakes to the final volume of our consistent set. However, we need a way to lower bound the final volume of our consistent set.

We define the \emph{width} of a polytope $K$ in dimension $i$ as $\mathrm{width}_i(K) = \max_{x,y \in K} |x_i - y_i|$. Note that the width can be efficiently computed using a linear program. We take advantage of the fact that the true valuation function takes values that are discretized to multiples of $\delta$. Hence, if at any point the width of our consistent set $C_{t}$ in some dimension $i$ is less than $\delta/2$, then we can exactly identify the $i^{th}$ coefficient of the consumer's valuation function. Note that if $\mathrm{Vol}(C_t) < \delta^n$, then there must be some dimension in which $\mathrm{width}_i(C_t) < \delta/2$. When we detect such a dimension, we fix that coefficient, and restart the algorithm by maintaining a consistent set in one fewer dimension.

Hence, at each \emph{epoch}, we maintain a set of consistent valuation functions restricted to those indices among those that are not yet fixed, and predict according to the composite of the valuation vectors sampled from this consistent set, together with the fixed indices. Every time we fix an index, a new epoch begins. The volume of the consistent set can never go below $\delta^n$ within an epoch, and since we fix an index at the end of every epoch, there can be at most $n$ such epochs.

The only computationally challenging step is sampling a point uniformly at random from the consistent set $C_t$, 
which can be done in polynomial time using the technique of \cite{DFK91}. 

\begin{varalgorithm}{ExogLearnVal($B$, $\delta$)}
\caption{}
\label{alg:learnvalexg}
\begin{algorithmic}
\State $\textrm{Fixed} = \emptyset, w_i = 0$ for all $i$. \Comment{Initialize the fixed coordinates, initially none}
\State $C_0 = \{z \in [0,1]^n \mid 0 \leq z_i \leq 1 \; \forall i\}$ \Comment{Initialize the set of consistent hypotheses}
\For{$t = 0$ to $T$}
\State Observe $p^t$
\State $z^t \leftarrow \texttt{sample}^1(C_t)$ \Comment{Sample a valuation uniformly from consistent set}
\State $v^t = (z^t_{-\textrm{Fixed}}, w_{\textrm{Fixed}})$ \Comment{Combine sampled and fixed coefficients}
\State Predict $\hat{x}^t \in \arg \max_{x \cdot p^t \leq B} x \cdot v^t$ \Comment{Predict according to sampled valuation}
\State $x_t \leftarrow \texttt{consumer}(p^t)$
\State $C_{t+1} = C_t$
\For{$i,j \not\in \textrm{Fixed} \mid x^t_i > x^t_j$} \Comment{Update Constraints}
  \State $C_{t+1} = C_{t+1} \cap \{z_i/p^t_i \geq z_j/p^t_j\}$
\EndFor
\For{$i \not\in \textrm{Fixed},j \in \textrm{Fixed} \mid x^t_i > x^t_j$} \Comment{Update Constraints}
  \State $C_{t+1} = C_{t+1} \cap \{z_i/p^t_i \geq w_j/p^t_j\}$
\EndFor
\For{$j \not\in \textrm{Fixed},i \in \textrm{Fixed} \mid x^t_i > x^t_j$} \Comment{Update Constraints}
  \State $C_{t+1} = C_{t+1} \cap \{w_i/p^t_i \geq z_j/p^t_j\}$
\EndFor
\If{There exists $i\not\in \textrm{Fixed}$ such that $\mathrm{width}_i(C_{t+1}) < \delta/2$} \Comment{Start a new Epoch}
  \For{Each $i\not\in \textrm{Fixed}$ such that $\mathrm{width}_i(C_{t+1}) < \delta/2$} \Comment{Fix each determined coordinate}
    \State $\mathrm{\textsc{Fix}}(i)$
  \EndFor
  $C_{t+1} = \{z \in [0,1]^{n-|\textrm{Fixed}|} \mid 0 \leq z_i \leq 1 \; \forall i\not\in \textrm{Fixed}\}$ \Comment{Re-initialize in the unfixed \\ \hfill coordinates}
\EndIf
\EndFor \\

\Procedure{Fix}{$i$}
\State $\textrm{Fixed} = \textrm{Fixed} \cup \{i\}$
\State $z \leftarrow \texttt{sample}(C_{t+1})$ \Comment{Sample a value for coordinate $i$}
\State $w_i = \texttt{round}^2(z_i/\delta) \cdot \delta$ \Comment{Round sampled value to nearest discrete value and fix it}\\
\EndProcedure

\State $^{1}$\texttt{sample} is any polynomial algorithm for sampling uniformly at random from a convex polytope.
\State $^{2}$\texttt{round}($x$) rounds to the nearest integer.

\end{algorithmic}
\end{varalgorithm}

\begin{theorem4}
\label{thm:exogenous}
Algorithm \emph{\texttt{ExogLearnVal}} runs in polynomial time per round, and with probability $1-\beta$, makes at most $O\left(n^2\log(1/\delta) + n\sqrt{\log(1/\beta)\log(1/\delta)}\right)$ mistakes over any sequence of adaptively chosen price vectors.
\end{theorem4}

\proof

We prove a bound on the number of mistakes made by the algorithm in a single epoch (recall that each epoch ends when a new coordinate is fixed). Since there are only $n$ coordinates, and hence at most $n$ epochs, our final mistake bound is at most $n$ times the mistake bound per epoch, giving the theorem.

Consider an epoch $i$ in which there remain $d \leq n$ unfixed coordinates. We will track the $d$-dimensional volume of the sets $C_t$ during this epoch. Let $S_i$ be the first stage of the epoch, and let $F_i$ be the final stage of the epoch. Note that $\mathrm{Vol}(C_{S_i}) = 1$ (because $C_{S_i}$ is always initialized to be the $d$-dimensional hypercube). Note also that for $S_i \leq t < F_i$, any hypothesis in $C_t$ that leads to an incorrect prediction of $x^t$ is eliminated from $C_{t+1}$. Hence, if $M_t$ is the indicator random variable specifying whether our algorithm makes a mistake at round $t$, it is also the indicator random variable specifying whether our algorithm sampled a hypothesis that will be eliminated in the next round.  Because we sample our hypothesis in round $t$ at random from $C_t$, we have:
$$\E[M_t] = 1 - \frac{\mathrm{Vol}(C_{t+1})}{\mathrm{Vol}(C_t)}$$

Note that we can write the volume of $C_{F_i}$ as a telescoping product:
$$\mathrm{Vol}(C_{F_i}) = \mathrm{Vol}(C_{S_i})\prod_{t=S_i}^{F_i-1}\frac{\mathrm{Vol}(C_{t+1})}{\mathrm{Vol}(C_{t})}=\prod_{t=S_i}^{F_i-1}\frac{\mathrm{Vol}(C_{t+1})}{\mathrm{Vol}(C_{t})}$$
Note also that before the end of an epoch, the width of $C_t$ in every coordinate is at least $\delta/2$. Hence we know $\mathrm{Vol}(C_{F_i}) \geq \delta^d \geq \delta^n$. Combining these facts, we can write:

$$\delta^n \leq \prod_{t=S_i}^{F_i-1}\frac{\mathrm{Vol}(C_{t+1})}{\mathrm{Vol}(C_{t})}
= \prod_{t=S_i}^{F_i-1} \left(1-\E[M_t]\right)
\le\prod_{t=S_i}^{F_i-1} \exp(-\E[M_t])
= \exp\left(-\E\left[\sum_{t=S_i}^{F_i-1} M_t\right]\right)$$

Solving for the expected number of mistakes made in a single epoch,
we find that:
$$\E\left[\sum_{t=S_i}^{F_i} M_t\right] \leq \E\left[\sum_{t=S_i}^{F_i-1} M_t\right] + 1 \leq 1 + n\ln\left(\frac{1}{\delta}\right) = O\left( n\ln\left(\frac{1}{\delta}\right)\right)$$

Now consider the expected number of mistakes made by our algorithm for its entire run, $t \in \{1,\ldots,T\}$. Since we can partition each time step into one of at most $n$ epochs, and have a bound on the expected number of mistakes in each epoch, by linearity of expectation, we have:

$$\E\left[\sum_{t=1}^{T} M_t\right] = \sum_{i=1}^n \E\left[\sum_{t=S_i}^{F_i} M_t\right] \leq n + n^2\ln\left(\frac{1}{\delta}\right) = O\left( n^2\ln\left(\frac{1}{\delta}\right)\right)$$

Note that each of the random variables $M_i$ is independent and bounded in $[0,1]$. We can therefore apply a multiplicative Chernoff bound. For any $\epsilon < 1$, we have:
$$\Pr\left[\sum_{t=1}^{T} M_t \geq (1+\epsilon)\E\left[\sum_{t=1}^{T} M_t\right]\right] \leq \exp\left(\frac{-\epsilon^2}{3}\E\left[\sum_{t=1}^{T} M_t\right]\right)$$

Setting the right hand side to be at most $\beta$, plugging in our bound on $\E\left[\sum_{t=1}^{T} M_t\right]$  and solving for $\epsilon$ allows us to take
$$\epsilon = O\left(\sqrt{\frac{\ln(1/\beta)}{n^2\ln(1/\delta)}}\right)$$
 Plugging this into the Chernoff bound proves the theorem.

\qed (Theorem 4)
